# Software Process Models and Analysis on Failure of Software Development Projects

Rupinder Kaur, Dr. Jyotsna Sengupta

**Abstract**— The software process model consists of a set of activities undertaken to design, develop and maintain software systems. A variety of software process models have been designed to structure, describe and prescribe the software development process. The software process models play a very important role in software development, so it forms the core of the software product. Software project failure is often devastating to an organization. Schedule slips, buggy releases and missing features can mean the end of the project or even financial ruin for a company. Oddly, there is disagreement over what it means for a project to fail. In this paper, discussion is done on current process models and analysis on failure of software development, which shows the need of new research.

—————————— ♦ ——————————

## 1 INTRODUCTION

Many software development models are described in the software engineering, which describes the states through which software evolves. The life-cycle focuses on the product, defining the state through which a product passes from when it starts to be built to when software enters into operations and finally retired [14]. A software process model is an abstract representation of the architecture, design or definition of the software process [15].

In software development, process models are implemented to manage various concerns associated with cost, time, and quality and changing requirements of client's etc. The particular life cycle model can significantly affect various concerns associated with a software product. If the process is weak, the end product certainly will suffer. Enough effort has been done in this field; still ever changing requirement during the development process for large software development is still not managed by software process models, which results in software projects not meeting their expectation in terms of functionality, cost and delivery schedule.

The reason of failure can be project team, suppliers, customers and other stakeholders, but the most common reasons for project failure are rooted in the project management process itself and the aligning of IT with organizational cultures [7].

The identified estimation mistakes, unclear project goals and objectives, and project requirement changing during the project are some key factors in project failures.

The remainder of the paper is organized as follows: Section 2 discuss existing models and techniques, Section 3 presents the analysis on failure of software developments, which shows that new research is required in this field. In Section 4 the conclusion is done.

## 2 BACKGROUND WORK

Developing and maintaining software systems involves a variety of highly interrelated activities. In order to manage these structured set of activities various models have been developed over the years with varying degree of success. These include Waterfall model, Iterative development, Prototyping, Spiral model, RAD. Each product can pass through different states, depending on the specific circumstances of each project and hence, there are different development models. For example, if the problem is well defined and well understood and user need is practically invariable, a short waterfall-like life cycle is likely to be sufficient. The Waterfall Model was widely used because it formalized certain elementary process control requirements. However, when we come up against a poorly defined and understood problem and a highly volatile user need, we can hardly expect to output a full requirements specification at the start. In this case, we have to opt for longer and more complex life cycles, like the Spiral Model [2].

Each cycle in Spiral Model addresses the development of the software product at a further level of detail. In the course of several papers, Boehm and his colleagues extended the spiral model to a variant called the Win–Win Spiral Model [2], [3], [4]. The win–win stakeholder approach is used to determine three critical project milestones that together anchor the development of the project: namely, life-cycle objectives, life-cycle architectures, and initial operational capability [1]. Prototyping Model helps to understand uncertain requirements but leads to false expectations and poorly designed system. A popular variation of the Prototyping Model is called Rapid Application Development (RAD). This model introduces firm time limits on each development phase and relies heavily on rapid application tools which allow for quick devel-opment [9]. Exploratory model use the prototyping as a technique for gathering requirements and was very simple in its construction but is limited with high level language for rapid development. This model works best in situations where few, or none, of the system or product requirements are known in detail ahead of time. This model is largely based on educated guesswork. This scheme is not particularly cost-effective and sometimes results in less-than-optimal systems, so it should be used only when no viable alternative seems to exist.

Agile process model give less stress on analysis and design. Implementation begins much early in the life cycle of the software development. This process model demands fixed time. Extreme Programming (XP) was created by Kent Beck during software development, and is based on iterative enhancement model. Like other agile software development, XP attempts to reduce the cost of change by having multiple short development cycles, rather than one long one. It only works on teams of twelve or fewer people [11]. Industrial Extreme Programming (IXP) was introduced as an evolution of XP. It is intended to bring the ability to work in large and distributed teams. It then supported flexible values [10]. There is not enough data to prove its usability.

These days, majority of the software development project involve some level of reuse of existing artifact like design or code modules. The component-based development (CBD) model incorporates many of the characteristics of the spiral model. It is evolutionary in nature, demanding an iterative approach to the creation of software [12]. The component-based development model leads to software reuse, and reusability provides software engineers with a number of measurable benefits. The unified software development process is representative of a number of component-based development models that have been proposed. Using a combination of iterative and incremental development, the unified process defines the function of the system by applying a scenario-based approach [8].The concentration is on object oriented development.

The evolution of software development Process Models has reflected the changing needs of software customers. As customers demanded faster results, more involvement in the development process and the inclusion of measures to determine risk and effectiveness and the methods for developing systems changed. Before requirements can be finalized we must understand the domain. According to Dines Bjorner, it is not possible to develop software without understanding its domain [6]. These rapid and numerous changes in the system development environment simultaneously spawned the development of more practical new Process Models and the demise of older models that were no longer useful [13].

## 3 ANALYSIS ON FAILURE OF SOFTWARE DEVELOPMENT

Software projects fail when they do not meet the criteria for success. Most of the IT projects run over budget or are terminated prematurely and those that reach completion often fall far short of meeting user expectations and business performance goals.

Here we discuss various reports on failure of software product, projected by Dan Galorath [5].There is several updates to the Standish "Chaos" reports. The 2004 report shows:

- Successful Projects: 29%
- Challenged Projects: 53%
- Failed Projects: 18%

Standish Findings By Year Updated for 2009:

|  | 1994 | 1996 | 1998 | 2000 | 2002 | 2004 | 2009 |
|---|---|---|---|---|---|---|---|
| Succeeded | 16% | 27% | 26% | 28% | 34% | 29% | 32% |
| Failed | 31% | 40% | 28% | 23% | 15% | 18% | 24% |
| Challenged | 53% | 33% | 46% | 49% | 51% | 53% | 44% |

**TCS (Tata Consultancy Services) 2007**

- 62% of organizations experienced IT projects that failed to meet their schedules.
- 49% suffered from budget overruns.
- 47% had higher-than-expected maintenance costs.
- 41% failed to deliver the expected business value and ROI.
- 33% file to perform against expectations.

**Avanade Research Report (2007)**

- 66% of failure due to system specification.
- 51% due requirement understanding.
- 49% due to technology selection.

**ESSU (European Service Strategy Unit) Research Report 2007**

- 57% of contracts experienced cost overruns.
- 33% of contracts suffered major delays.
- 30% of contracts were terminated.
- 12.5% of Strategic Service Delivery Partnerships have failed.

**KPMG Survey (2008)**

On average, about 70 % of all IT-related projects fail to meet their objectives.

**From Bob Lawhorn presentation on software failure (March 2010)**

- Poorly defined applications (miscommunication between business and IT) contribute to a 66% project failure rate, costing U.S. businesses at least $30 billion every year (Forrester Research)
- 60% – 80% of project failures can be attributed directly to poor requirements gathering, analysis, and management (Meta Group)
- 50% are rolled back out of production (Gartner)
- 40% of problems are found by end users (Gartner)
- 25% – 40% of all spending on projects is wasted as a result of re-work (Carnegie Mellon)
- Up to 80% of budgets are consumed fixing self-inflicted problems (Dynamic Markets Limited 2007 Study)

The three major key factor of project success are delivered on time, on or under budget, the system works as needed. Few projects achieve all three. Many more are delivered which fail on one or more of these criteria, and a substantial number are cancelled having failed badly. They are number of software projects that succeeded or failed. So are the key factors for success of a project is based on only these three criteria? They is no one factor that cause the failure of project, a number of factors are involved. Some of the most vital reasons for failure are as follows:

**3.1 Extracting Requirements**

Extracting requirements of a desired software product is the first task in creating it. Sometimes the goal of a project may be only partially clear due to a poor requirement gathering in the definition stage of a project. Many projects have high level, vague and generally unhelpful requirements. This leads the developers having no input from the user and build what they believe is required, without having any real knowledge of the business for which the project is being developed. Inevitably when the system is delivered, user declares, it does not do what they needed it to. Defining clear requirements for a project can take time and lots of communication, but sometimes goals and objectives might be unclear because project sponsors lack the experience to describe what they really require. User should know what they require from the project and be able to specify it clearly. However as user is non-IT specialist, developer must extracting requirements from the user through his/her skills and experience in software engineering.

**3.2 Lack of User Involvement**

The research companies and academic institution has focused on the lack of executive support and user involvement as two main difficulties in managing IT projects. Lack of user involvement has proved fatal for many projects. Without user involvement nobody in the business feels committed to a system and can even be hostile to it. One of the criteria, of the software project success depends on user involved from the start of the project and continuously throughout the development. This requires time and effort from user end, which is often not done as finding time for a new project is not high on their priorities. User needs to continuously support the project. The developer must involve the user, as it helps in requirements elicitation and delivering all functionality of the project.

**3.3 Team Size**

Proper team size is essential in software development project. Basically there are three different project team sizes: small team of 10 or fewer people for small project, medium size team of 11 to 25 people for medium project and large team of 26 or more for large project. Small group of team results in good communication and tends to be very flexible over large group of teams. It is easy to call meetings and get instant feedback. Projects sometimes fail due to improper communication.

**3.4 Time Dimension**

Time dimension is important in software development, it is beneficial to deliver project on given time schedule. So given time should be appropriately well thought-out. The time on task is the time the task will take to complete without interruptions, whereas duration is the time the task actually take to complete including interruptions. Using the time on task to estimate schedule is one of the common mistakes made by project managers. Long timescales for a project, led the project to fail and no longer is required by an organisation. The key recommendation is that project time dimension should be short.

**3.5 Fixed Controller**

It is not realistic to expect no change in requirements while a system is being built. As an environment constraints and client requirement keep on changing, developer must follow component driven approach while building the system. The new requirements or modifications can be taken care separately till these components match with the current development process. However uncontrolled changes play havoc with the system under development and have cause many project failures.

### 3.6 Testing

A primary purpose of testing is to detect software failures so that defects may be discovered and corrected. Developer do testing of software product during development but eventually the user must run the acceptance testing to see if the system meets the business requirement. Often acceptance testing fails to catch many faults before system goes live, as it may be due to unplanned testing, inadequately trained user who do not known the purpose of testing and inadequate time to perform testing as the project is late.

### 3.7 Poor Quality management

Periodic quality evaluation and appropriate prevention & removal measures are mandatory if the quality of the deliverable needs to be as desired. Examples of defect removal activities include requirements review, design review, code review and different kinds of testing.

## 4 CONCLUSION

This paper makes an attempt to study variety of software process models and analysis various issues in software development projects. Discussion is done on various reports, which exhibit the failure of software product. Projects run over budget or are terminated prematurely and those that reach completion often fall far short of meeting user expectations and business functionalities. Few vital factors are discussed that cause the failure of projects. These factors are not the only one that affect the success or failure of a project, but they are among the near or the top of the list. This study shows the need to develop a new approach, model or techniques to resolve the major issues of software development.